\documentclass{ws-procs9x6}

\def\etal{{\it et al}}

\def\NPB#1#2#3{Nucl. Phys. B \textbf{#1},  #3 (#2)}
\def\PLB#1#2#3{Phys. Lett. B \textbf{#1},  #3 (#2)}
\def\PLA#1#2#3{Phys. Lett. A \textbf{#1},  #3 (#2)}
\def\PRD#1#2#3{Phys. Rev. D  \textbf{#1},  #3 (#2)}

\def\PRL#1#2#3{Phys. Rev. Lett. \textbf{#1},  #3 (#2)}
\def\PRT#1#2#3{Phys. Rep. {\textbf#1},  #3 (#2)}

\def\IJMP#1#2#3{Int. J. Mod. Phys. A {\textbf #1}, #3 (#2)}

\def\JHEP#1#2#3{JHEP {\textbf #1}, #3 (#2)} 
\begin{document}

\title{
\rightline{OUTP--03--34P}
\rightline{\tt hep-th/0312167}
\rightline{December 2003}
\rightline{}
Anthropics Versus Determinism \\in Quantum Gravity}



\address{Dedicated to the memory of Ian Kogan}

\author{Alon E. Faraggi\footnote{to appear in the proceedings of
the second International Conference on String Phenomenology, 
29/7-4/8 2003, Durham UK}
}

\address{Theoretical Physics, University of Oxford, Oxford, OX1 3NP, England}


\maketitle

\abstracts{
Recently the multitude of vacua in string theory have led some authors
to advocate the anthropic principle as a possible resolution for the
contrived set of parameters that seem to govern our universe. I suggest
that string theories should be viewed as effective theories, 
and hence of limited utility rather than as ``theories of everything''.
I propose that quantum gravity should admit a form of determinism
and that the self--dual points under phase--space duality should
play a prominent role in the vacuum selection principle. 
}

\section{Introduction}

String theory is in a precarious state of affairs. On the one hand
the theory clearly shows great promise in providing a consistent
theory of perturbative quantum gravity, while at the same time
accommodating the gauge and matter structures that are observed 
experimentally. On the other hand the multitude of string vacua
hinders the prospects of extracting unique predictions from the 
theory, and has led some authors to advocate the anthropic principle
as a resolution for understanding the contrived set of parameters
that seem to govern our world \cite{anthropics}.

Part of the difficulty stem from the desire to regard 
string theory as a ``theory of everything". However, 
even if string theory does lead to the final theory, 
a more cautious approach towards its utilization seems prudent.
In this respect, while the theory may offer a glimpse into 
some of the building blocks of quantum gravity and its 
relation to observable phenomena, the eventual formalism
of quantum gravity and its physical picture may be 
entirely different from our contemporary perception of
string theory. The important question is whether string theory
reveals some of the basic properties of the yet unknown quantum 
gravity theory. Our task is therefore to try to extract from
string theory what those properties may be. 

In this regard we may view string theories as providing
consistent approaches to the synthesis of quantum mechanics
and gravity. What lesson should we then extract from the existence
of a multitude of string vacua. One possible conclusion is that this
is a mere reflection of pushing the probability
interpretation of quantum mechanics to its logical conclusion. 
Namely, the different string theories
are the inevitable consequence of applying the probability interpretation
of quantum mechanics to the space--time arena. This is of course not new,
but a mere example, albeit perhaps
in a more advanced framework, of the multiverse
interpretation of quantum gravity. Pursuing this interpretation
it therefore seems to me that the anthropic principle is the unavoidable 
result of pushing the probability interpretation of 
quantum mechanics to its logical conclusion. 

What may therefore be at hand is the inevitable clash between
the basic physical hypothesis that underly quantum mechanics and
classical general relativity. String theory adopts the probability
interpretation of quantum mechanics from the outset.
String theory may of course be very useful in allowing 
a perturbative scheme to quantum gravity, and hence 
enabling detailed approximate calculations to be 
performed. However, we should not overlook the possibility
that the probability interpretation of quantum mechanics
is a derived property rather than a fundamental property
of quantum gravity.

In this article I would like to suggest that there exist
an alternative possibility. I argue that the evidence
suggests that quantum gravity should admit a form of determinism.
In this context it is proposed that the self--dual 
points under phase--space duality should play an important role 
in the vacuum selection mechanism and in the eventual formulation
of quantum gravity. This evidence arises from two considerations.
The first is the general expectation
that the self--dual points under T--duality in toroidal compactifications
extremizes the potential for the moduli of the compactified dimensions
in superstring theories.
It is then extremely intriguing that the most realistic string models
constructed to date, the so-called free fermionic models, are in fact
built in the vicinity of the self--dual point.
The second piece of evidence
arises from the relation of self--dual points under phase--space
duality to the vacuum state, in a modified Hamilton--Jacobi approach
to quantum mechanics. These two completely disparate considerations,
in my view, provide two crucial hints in the pursuit of quantum 
gravity. That the fundamental duality is phase--space duality,
and that the self--dual points under this duality play a pivotal
role in the vacuum selection.

\section{T--duality}

String theory exhibits various forms of dualities, {\it i.e.} 
relation between different theories at large and small radii
of the compactified manifold and at strong and weak coupling. 
The first type is the T--duality \cite{tduality}. Consider a point particle 
moving on a compactified dimension $X$, which obeys the condition
$X\sim X +2\pi R m$. Single valuedness of the wave function of the 
point particle $\Psi\sim {\rm Exp}(iP X )$ implies that the momenta
in the compact direction is quantized $P={m\over R}$ with $m\in Z$. 
Now consider a string moving in the compactified direction.
In this case the string can wrap around the compactified dimension
and produce stable winding modes. Hence the left and right--moving
momenta in the case of the closed string are given by
$$P_{L,R}={m\over R}\pm {{nR}\over \alpha^\prime}$$
and the mass of the string states is given by
$${\rm mass}^2=\left({n\over R}\right)^2+\left({{m R}\over
\alpha^\prime}\right)^2$$
this is invariant under exchange of large and small radius together 
with the exchange of winding and momentum modes, {\it i.e.}
$${1\over R}\leftrightarrow {R\over\alpha^\prime}~~~{\rm with}~~~
m\leftrightarrow n$$
and is an exact symmetry in string perturbation theory. Furthermore, 
there exist the self--dual point, 
$$ R={\alpha^\prime\over R},$$
which is the symmetry point 
under T--duality. In the case of the supersymmetric string on a 
compactified coordinate the T--duality operation interchanges 
\begin{eqnarray}
{\rm type ~IIA} & ~\Leftrightarrow~ & {\rm type~ IIB}\\
{\rm Heterotic~} SO(32) & ~\Leftrightarrow~ & {\rm Heterotic}~ E_8\times E_8
\end{eqnarray}
Now, all this
is of course well known since the late 80's. However, the following point
is not well appreciated. It is also well known that for specific
values of its  radius, the compactified coordinate can be realized 
as specific rational conformal field theories propagating on
the string world--sheet. In particular, there exist such a value for
which a compactified coordinate can be represented in terms
of two free Majorana--Weyl fermions. It so happens that, in some
normalization, the self--dual point is at $R=1/\sqrt2$ whereas the
free fermionic point is at $R=1$. Hence, the two points do not overlap
and the free fermionic point does not coincide with the self--dual point \cite{ginsparg}. 
However, this is merely an artifact of the fact that we have been talking
here about bosonic string. In the case of the supersymmetric string the 
two points, in fact, do coincide. This is a remarkable observation for
the following reason. While we do not yet know at what value the compactified
coordinate are fixed, naively we would expect that they are stabilized
around a symmetry point or at infinity. The self--dual point under T--duality
is precisely such a symmetry point. Hence, near the self--dual point,
which is the symmetry point under T--duality and around which
we may expect that the compactified dimensions stabilize, we can represent
the compact dimension as a pair of free Majorana--Weyl fermions
propagating on the string world--sheet. Of course, the real picture 
may be much more complicated. But as a first approximation this is the 
naive expectation, based on the symmetry properties of string theory.

\section{Realistic string models}\label{rsm}

The class of three generation free fermionic models is therefore 
constructed precisely in the vicinity of the self--dual point under 
T--duality! I find this to be an extremely intriguing coincidence!
I elaborate briefly here on the properties of these models
and why they may be considered the most realistic models
constructed to date.

{}From the Standard Model data we may hypothesis that the string vacuum
should possess two key properties.
The existence of three generations and their 
embedding in $SO(10)$ representations.
The only perturbative string theory
that preserves the $SO(10)$ embedding is the heterotic string, because
this is the only one that produces the chiral 16 representation
of $SO(10)$ in the perturbative spectrum. 
A class of heterotic string models that accommodate these
two key ingredients of the Standard Model spectrum are the 
so called free fermionic models. As already discussed above
the free fermionic point in the moduli space of superstring theories
coincides with the self--dual point. The other key property of the 
free fermionic models is their relation to $Z_2\times Z_2$
orbifold compactification. While the space of possible 
string compactifications may be beyond count, it seems to me
that any model, or class of models, that exhibit realistic 
properties deserve to be studied in depth. 

The structure of the $Z_2\times Z_2$ orbifold naturally
correlates the existence of three generation
with the underlying geometry. This arises due to the fact that 
the $Z_2\times Z_2$ orbifold has exactly three twisted sectors. 
Each of the light chiral generations then arises from a distinct 
twisted sector. Hence, in these models the existence of three generations
in nature is seen to arise due to the fact that we are dividing
a six dimensional compactified manifold into factors of 2. In simplified
terms, three generations is an artifact of 
$
{6\over 2}~~=~~1~+~1~+~1~.~
$
One may further ask whether there is a reason that the $Z_2$ 
orbifold would be preferred versus higher orbifolds.
Previously we argued that the free fermionic point 
coincides with the self--dual point under $T$--duality,
which is where we would naively expect the compactified 
dimensions to stabilize. The special property of the 
$Z_2$ orbifold that sets it apart from higher orbifolds,
is the fact that the $Z_2$ orbifold is the only one that acts
on the coordinates as real coordinates rather than complex 
coordinates. Furthermore, a classification of the $Z_2\times Z_2$
orbifold with symmetric shifts on the internal tori reveals that the
three generation models in this class 
are not obtained solely with symmetric 
shifts on complex tori \cite{fknr}. The three generation models
necessarily employ an asymmetric shift on the internal coordinates.
The necessity to incorporate an asymmetric shift
in the reduction to three generations, has profound implications for the
issues of moduli stabilization and vacuum selection. The reason
being that it can only be implemented at enhanced symmetry
points in the moduli space. In this context we envision again that the
self--dual point under T--duality plays a special role. In the 
context of nonperturbative dualities the dilaton and
moduli are interchanged, with potentially important
implications for the problem of dilaton stabilization.

To summarize this section. The argument here is that T--duality
is the pivotal property of string theory in trying to understand
the vacuum selection mechanism. In this context the self--dual points
may play an important role. It is then extremely intriguing that
precisely in the vicinity of the self--dual point there exist 
a class of models that capture the two main characteristics
of the Standard Model. The existence of three generations
together with their $SO(1O)$ embedding.

\section{Phase--space duality}

Duality and self--duality also play a key role in the recent
formulation of quantum mechanics from an equivalence
postulate \cite{fm}.
An important facet of this formalism is the
phase--space duality, which is manifested due to the
involutive nature of the Legendre transformation.
In the Hamilton--Jacobi formalism of classical mechanics
the phase--space variables are related by Hamilton's generating
function $p=\partial_q{S}_0(q)$.
One then obtains the dual Legendre transformations \cite{fm},
$$
{ S}_0=p\partial_p{ T}_0-{ T}_0
$$
and
$$
{ T}_0=q\partial_q{ S}_0-{S}_0,
$$  
where ${ T}_0(p)$ is a new generating function
defined by $q=\partial_p{ T}_0$.
Two points are important to note. The first is that 
because the Legendre
transformation is not defined for linear functions, {\it i.e.} for
physical systems with ${ S}_0=A q+B$, implies that the
Legendre duality fails for the free system and for the
free system with vanishing energy. 
The second is that we can associate a second order differential 
equation with each Legendre transformation \cite{fm}. 
There exist then a set of solutions, labeled by $pq=\gamma$,
where $\gamma$ is a constant to be determined, which are simultaneous
solutions of the two sets of differential equations. These are the 
self dual states under the phase--space duality. 

\section{The quantum equivalence postulate}

The Legendre phase--space
duality and its breakdown for the free system are intimately
related to the equivalence postulate, which states
that all physical systems labeled by the function
${ W}(q)=V(q)-E$, can be connected by a coordinate
transformation, $q^a\rightarrow q^b=q^b(q^a)$, defined
by ${ S}_0^b(q^b)={ S}_0^a(q^a)$.
This postulate implies that there
always exist a coordinate transformation connecting  
any state to the state ${ W}^0(q^0)=0$. Inversely, this means
that any physical state can be reached from the
state ${ W}^0(q^0)$ by a coordinate transformation.
This postulate cannot be consistent
with classical mechanics. The reason being that in Classical
Mechanics (CM) the state ${ W}^0(q^0)\equiv0$ remains a fixed
point under coordinate transformations. Thus, in CM it
is not possible to generate all states by a coordinate
transformation from the trivial state. Consistency of the
equivalence postulate implies the modification of CM,
which is analyzed by a adding a still unknown function
$Q$ to the Classical Hamilton--Jacobi Equation (CHJE).
Consistency of the equivalence postulate fixes the
transformation properties for ${ W}(q)$,
$$
{ W}^v(q^v)=
 \left(\partial_{q^v}q^a\right)^2{ W}^a(q^a)+(q^a;q^v),
$$
and for $Q(q)$,
$$
 Q^v(q^v)=\left(\partial_{q^v}q^a\right)^2Q^a(q^a)-(q^a;q^v),
$$
which fixes the cocycle condition for the inhomogeneous term
$$
(q^a;q^c)=\left(\partial_{q^c}q^b\right)^2[(q^a;q^b)-(q^c;q^b)].
$$
The cocycle condition is invariant under M\"obius transformations
and fixes the functional form of the inhomogeneous term.
The cocycle condition is generalizable to higher, Euclidean or
Minkowski, dimensions,
where the Jacobian of the coordinate transformation extends
to the ratio of momenta in the transformed and original systems\cite{fm}.
 
The identity   
$$
({\partial_q{ S}_0})^2=
\hbar^2/2\left(\{\exp(i2{ S}_0/\hbar,q)\}-\{{ S}_0,q\}\right),
$$
which embodies the equivalence postulate,
leads to the Schr\"odinger equation.
Making the identification
$$
{ W}(q)= V(q) - E = -{\hbar^2/{4m}}\{{\rm e}^{(i2{S}_0/\hbar)},q\},
$$
and
$$
{Q}(q)=  {\hbar^2/{4m}}\{{ S}_0,q\},
$$
we have that   
${ S}_0$ is solution of the Quantum Stationary
Hamilton--Jacobi Equation (QSHJE),
$$
({1/{2m}})\left({{\partial_q S}_0}\right)^2+
V(q)-E+({\hbar^2/{4m}})\{{ S}_0,q\}=0,
$$
where $\{,\}$ denotes the Schwarzian derivative.
{}From the identity we deduce that the trivializing
map is given by $q\rightarrow {\tilde q}=\psi^D/\psi$,
where $\psi^D$ and $\psi$ are the two linearly independent
solutions of the corresponding Schr\"odinger equation \cite{fm}.
We see that the consistency of the equivalence postulate
forces the appearance of quantum mechanics and
of $\hbar$ as a covariantizing parameter.

\section{The role of the self--dual states}
The remarkable property of the QSHJE, which distinguishes
it from the classical case, is that it admits non--trivial solution
also for the trivial state, ${ W}(q)\equiv0$.
In fact the QSHJE implies that ${ S}_0=constant$ is
not an allowed solution. The fundamental characteristic
of quantum mechanics in this approach is that ${ S}_0\ne Aq+B$.
Rather, the solution for the ground state, with $V(q)=0$ and $E=0$,
is given by
$$
{ S}_0=i\hbar/2\ln q,
$$
up to M\"obius transformations. Remarkably, this quantum
ground state solution coincides with the self--dual state
of the Legendre phase--space transformation and its dual.
Thus, we have that the quantum self--dual state plays a pivotal
role in ensuring both the consistency of the equivalence
postulate and definability of the Legendre phase--space  
duality for all physical states. The association of the
self--dual state and the physical state with $V(q)=0$ and
$E=0$ provides a hint that the equivalence postulate   
and Legendre phase--space duality may shed new light
on the nature of the vacuum.

Two additional properties of the formalism are important
to note. The first is the existence of an intrinsic
length scale, which is strictly related to the phase--space
duality. Considering solutions of the basic Schr\"odinger equation
with $V(q)=E=0$, {\it i.e.}
$\partial_q^2\psi^0=0$, one has to introduce a length to consider linear 
combinations of $\psi^{D^0}=q^0$ and $\psi^0=1$. 
Since in this case ${W}\equiv0$, there is no scale in
the Schr\"odinger problem, and a universal length scale is enforced.
The second is the existence of equivalence classes of the wave--function.
As the QHJE is a third--order differential equation whereas the Schr\"odinger
equation is a second order one, more initial conditions
are needed to be specified in the case of the QHJE.
It follows that the wave function remains
invariant under suitable transformations of $\delta=\{\alpha, \ell\}$,
corresponding to different deterministic trajectories.
The implication is that there
are hidden variables which depend of the Planck length and
that these can suitably change without affecting the 
wave--function. The probabilistic nature of the wave--function
may therefore arise due to our ignorance of the underlying
Planck scale physics.

\section{Conclusions}

It is proposed that phase--space duality is the guiding 
property in trying to formulate quantum gravity. In this
respect T--duality is a key property of string theory.
We can think of T-duality as a phase--space duality
in the sense that we are exchanging momenta and winding
modes in compact space. We can turn the table around
and say that the key feature of string theory is that
it preserves the phase--space duality in the compact
space. Namely, prior to compactification the wave--function 
of a point particle $\Psi\sim{\rm Exp}(i P X)$ is invariant under
$p\leftrightarrow x$. However, in the ordinary Kaluza--Klein
compactification this invariance is lost due to the 
quantization of the momentum modes. String theory restores
this invariance by introducing the winding modes.
It is further argued that the self--dual points
under phase--space duality are intimately
connected to the choice of the vacuum. The evidence
for this arises from the phenomenological success of the
free fermionic models that are constructed in the vicinity 
of the self--dual point, as well as from the formal derivation
of quantum mechanics from phase--space duality and the
equivalence postulate. 

\section*{Acknowledgments}

This  work was supported in part by the PPARC and by the Royal Society.

\end{document}